\documentclass[11pt]{article}
\usepackage{amssymb}
\usepackage{amsmath}
\usepackage{graphicx}
\usepackage{cancel}

\numberwithin{equation}{section}

\begin{document}
\baselineskip=15pt
\begin{titlepage}
\begin{flushright}
{\small OU-HET 622/2009}\\[-1mm]%
{\small UT-HET 023}\\[-1mm]%
{\small KUNS-2186}%
\end{flushright}

\begin{center}
\vspace*{13mm}

{\Large\bf Observable Seesaw and its Collider Signatures}\\
\vspace*{10mm}

Naoyuki Haba$^1$, Shigeki Matsumoto$^2$, and Koichi Yoshioka$^3$\\
\vspace{4mm}

${}^1${\it Department of Physics, Graduate School of Science, 
Osaka University,\\[-.5mm]%
Toyonaka, Osaka 560-0043, Japan}\\[.5mm]%
${}^2${\it Department of Physics, University of Toyama, 
Toyama 930-8555, Japan}\\[.5mm]%
${}^3${\it Department of Physics, Kyoto University, 
Kyoto 606-8502, Japan}
\vspace*{10mm}

\begin{abstract}\noindent%
We discuss the scenario with TeV-scale right-handed neutrinos, which
are accessible at future colliders, while holding down tiny
seesaw-induced masses and sizable couplings to the standard-model
particles. The signal with tri-lepton final states and large missing
transverse energy is appropriate for studying collider signatures of
the scenario with extra spatial dimensions. We show that the LHC
experiment generally has a potential to discover the signs of extra
dimensions and the origin of small neutrino masses.
\end{abstract}

\end{center}
\end{titlepage}

\newpage
\section{Introduction}

The recent neutrino oscillation experiments have been revealing the
detailed structure of leptonic flavors~\cite{review,analysis}. The
neutrino property, in particular the tiny mass scale is one of the
most important experimental clues to find the new physics beyond the
standard model~(SM)\@. The seesaw mechanism naturally leads to small
neutrino masses by the integration of new heavy particles which
interact with the ordinary neutrinos. The introduction of heavy
right-handed neutrinos~\cite{seesaw} implies the intermediate mass
scale of such states to have light Majorana masses of order eV, and
these heavy states are almost decoupled in the low-energy effective
theory. Alternatively, TeV-scale right-handed neutrinos could also be
possible, which in turn means tiny orders of couplings to the SM
sector and their signs cannot be observed in near future TeV-scale
particle experiments such as the Large Hadron Collider (LHC).

The SM neutrinos have tiny masses due to a slight violation of the
lepton number. This fact implies that the events with same-sign
di-lepton final states~\cite{dileptons} are too rare to be
observed. In this letter, we focus on the lepton number preserving 
processes, in particular, the tri-lepton signals with large missing
transverse energy, $pp\to\ell^\pm\ell^\mp\ell^\pm\nu(\bar\nu)$. These
processes would be rather effectively detected at the LHC because only
small fraction of SM processes contributes to the background against
the signals.

As a simple example of observable seesaw theory, we consider a
five-dimensional extension of the SM with right-handed neutrinos,
where all SM fields are confined in the four-dimensional world, while
right-handed neutrinos propagate in the whole extra-dimensional
space~\cite{DDG}-\cite{neuExD2}. We will discuss an explicit framework
which provides the situation that TeV-scale right-handed neutrinos
generate tiny scale of seesaw-induced neutrino masses and
simultaneously have sizable interactions to the SM leptons and gauge
bosons. The scenario does not rely on any particular generation
structure of mass matrices and is available for one-generation
case. For such TeV-scale particles with large couplings to the SM
sector, the LHC experiment generally has the potential to find the
signals of extra dimensions and the origin of small neutrino masses.

\bigskip

\section{Observable Seesaw}

Let us consider a five-dimensional theory where the extra space is
compactified on the $S^1/Z_2$ orbifold with the radius $R$. The
SM fields are confined on the four-dimensional boundary
at $x^5=0$. Besides the gravity, only SM gauge singlets can propagate
in the bulk not to violate the charge
conservation~\cite{DDG,neuExD}. The gauge-singlet Dirac 
fermions ${\cal N}_i$ ($i=1,2,3$) are introduced in the bulk which
contain the right-handed neutrinos and their chiral partners. The 
Lagrangian up to the quadratic order of spinor fields is given by
\begin{eqnarray}
  {\cal L} \;=\; i\overline{\cal N}D\hspace{-2.5mm}/\,{\cal N}
  -\frac{1}{2}\big[\,\overline{{\cal N}^c}
  (M_v+M_a\gamma_5){\cal N}+\text{h.c.}\big].
\end{eqnarray}
The conjugated spinor is defined 
as ${\cal N}^c=\gamma_3\gamma_1\overline{\cal N}{}^{\rm t}$ such that
it is Lorentz covariant in five dimensions. It is straightforward to
write a bulk Dirac mass for ${\cal N}_i$ if introducing a $Z_2$-odd
function which originates from some field expectation value. The bulk
mass parameters $M_v$ and $M_a$ are $Z_2$ parity even and could depend
on the extra dimensional coordinate $x^5$ which comes from the
delta-function dependence (resulting in localized mass terms) and/or
the background geometry such as the warp factor in AdS$_5$. We also
introduce the mass terms between bulk and boundary fields:
\begin{eqnarray}
  {\cal L}_m \;=\; -\big(\overline{\cal N} mL+
  \overline{{\cal N}^c}m^cL\big)\delta(x^5) +{\rm h.c.},
  \label{boundary}
\end{eqnarray}
where $m$ and $m^c$ denote the mass parameters after the electroweak
symmetry breaking. The boundary spinors $L_i$ ($i=1,2,3$) contain the
left-handed neutrinos $\nu_i$, namely, given in the 4-component
notation $L_i=\genfrac{(}{)}{0pt}{1}{0}{\,\nu_i\,}$. The $Z_2$ parity
implies that either component in a Dirac fermion ${\cal N}$ vanishes at
the boundary ($x^5=0$) and therefore either of $m$ and $m^c$ becomes
irrelevant.\footnote{The exception is the generation-dependent $Z_2$
parity assignment on bulk fermions~\cite{HWY}. We do not consider such
a possibility in this paper.} In the following we assign the 
even $Z_2$ parity to the upper (right-handed) component of bulk
fermions, i.e.\ ${\cal N}(-x^5)=\gamma_5{\cal N}(x^5)$, and will 
drop the $m^c$ term.

With a set of boundary conditions, the bulk fermions ${\cal N}_i$ are
expanded by Kaluza-Klein (KK) modes with their kinetic terms being
properly normalized
\begin{eqnarray}
  {\cal N}(x,x^5) \;=\; \Bigg(\begin{array}{l}
    \sum\limits_n \chi^n_R(x^5)N_R^n(x) \\[1mm]
    \sum\limits_n \chi^n_L(x^5)N_L^n(x)
  \end{array} \Bigg).
\end{eqnarray}
The wavefunctions $\chi_{R,L}^n$ are generally matrix-valued in the
generation space and we have omitted the generation indices for
notational simplicity. After integrating over the fifth dimension, we
obtain the neutrino mass matrix in four-dimensional effective
theory. Neutrinos are composed of the boundary ones and the KK 
modes $(\nu,\epsilon N_R^{\,0\,*},\epsilon N_R^{\,1\,*},
N_L^{\,1},\cdots)\equiv(\nu,N)$. The zero modes of the left-handed
components have been extracted according to the boundary
condition. The neutrino mass matrix for $(\nu,N)$ is given by
\begin{eqnarray}
\renewcommand{\arraystretch}{1.15}
\qquad \left(%
\begin{array}{c|cccc}
& \,m_0^{\rm t} & \,m_1^{\rm t} & 0 & \cdots \\ \hline
m_0 & M_{R_{00}}^* & M_{R_{01}}^* & M_{K_{01}}^{} 
& \cdots \\[1mm]
m_1 & M_{R_{10}}^* & M_{R_{11}}^* & M_{K_{11}}^{} & \cdots \\[1mm]
0 & M_{K_{10}}^{\rm t} & M_{K_{11}}^{\rm t} 
& M_{L_{11}}^{} & \cdots \\[1mm]
\vdots & \vdots & \vdots & \vdots & \ddots
\end{array}\right)
\;\;\equiv\;\; -\!\left(
\begin{array}{c|ccc}
  & & M_D^{\rm t} & \\ \hline
  & & & \\
  \!\!M_D & ~ & M_N  & \\
  & & &
\end{array}\right),
\end{eqnarray}
where the boundary Dirac masses $m_n$, the KK masses $M_K$, and the
Majorana masses $M_{R,L}$ are
\begin{alignat}{2}
  m_n \;&= \, \chi^n_R\!{}^\dagger(0)m\,,& \qquad
  M_{R_{mn}} &= \int_{-\pi R}^{\pi R}\!\!dx^5\,
  (\chi^m_R)^{\rm t} (M_a+M_v) \chi^n_R\,, \nonumber \\
  M_{K_{mn}} &= \int_{-\pi R}^{\pi R}\!\!dx^5\,
  (\chi^m_R)^\dagger\partial_5\chi^n_L\,,& \qquad
  M_{L_{mn}} &= \int_{-\pi R}^{\pi R}\!\!dx^5\,
  (\chi^m_L)^{\rm t} (M_a-M_v) \chi^n_L\,.
\end{alignat}
It is noticed that $M_{K_{mn}}$ becomes proportional 
to $\delta_{mn}$ if $\chi_{R,L}^n$ are the eigenfunctions of the bulk
equations of motion, and $M_{R,L_{mn}}$ also becomes proportional 
to $\delta_{mn}$ if the bulk mass parameters $M_a$, $M_v$ are
independent of the coordinate $x^5$. 

We further implement the seesaw operation 
assuming ${\cal O}(m_n)\ll{\cal O}(M_K)$ or ${\cal O}(M_{L,R})$ and
find the induced Majorana mass matrix for three-generations light
neutrinos
\begin{eqnarray}
  M_\nu \;=\; M_D^{\text{t}}M_N^{-1}M_D^{}.
\end{eqnarray}
It is useful for later discussion of collider phenomenology to write
down the electroweak Lagrangian in the basis where all the mass
matrices are generation diagonalized. The interactions to the
electroweak gauge bosons are given in this mass eigenstate 
basis $(\nu_d,N_d)$ as follows:
\begin{eqnarray}
  {\cal L}_g = \frac{g}{\sqrt{2}}\Big[W_\mu^\dagger e^\dagger\sigma^\mu
  U_{\rm MNS} \big(\nu_d+VN_d\big) +\text{h.c.}\Big]
  +\frac{g}{2\cos\theta_W}Z_\mu\big(\nu_d^\dagger+
  N_d^\dagger V^\dagger\big)\sigma^\mu\big(\nu_d+VN_d\big),\;
\end{eqnarray}
where $W_\mu$ and $Z_\mu$ are the electroweak gauge bosons and $g$ is
the $SU(2)_{\rm weak}$ gauge coupling constant. The 2-component 
spinors $\nu_d$ are three light neutrinos for which the seesaw-induced
mass matrix $M_\nu$ is diagonalized
\begin{eqnarray}
  M_\nu \;=\; U_\nu^*\,M_\nu^d\,U_\nu^\dagger,  \qquad
  U_\nu\,\nu_d \;=\; \nu-M_D^\dagger M_N^{-1\,*}N,
\end{eqnarray}
and $N_d$ denote the infinite number of neutrino KK modes for which
the bulk mass matrix $M_N$ is diagonalized in the generation and KK
spaces by a unitary matrix $U_N\,$:
\begin{eqnarray}
  M_N \,=\; U_N^*\,M_N^d\,U_N^\dagger,  \qquad
  U_N N_d\ \,=\; N+M_N^{-1}M_D^{}\,\nu.
\end{eqnarray}
The lepton mixing matrix measured in the neutrino oscillation
experiments is given by $U_{\rm MNS}=U_e^\dagger U_\nu$ where $U_e$ is
the left-handed rotation matrix for diagonalizing the charged-lepton
Dirac masses. It is interesting to find that the model-dependent parts
of electroweak gauge vertices are governed by a single 
matrix $V$ which is defined as
\begin{eqnarray}
  V \;=\; U_\nu^\dagger M_D^\dagger M_N^{-1\,*}U_N.
\end{eqnarray}
When one works in the basis where the charged-lepton sector is flavor
diagonalized, $U_\nu$ is fixed by the neutrino oscillation matrix.

The neutrinos also have the interaction to the electroweak doublet
Higgs $H$ in four dimensions. The boundary Dirac 
mass~\eqref{boundary} comes from the Yukawa coupling
\begin{eqnarray}
  {\cal L}_h \;=\; -\big(y\overline{\cal N}LH^\dagger
  +\text{h.c.}\big)\delta(x^5).
\end{eqnarray}
The doublet Higgs $H$ has a non-vanishing expectation value $v$ and
its fluctuation $h(x)$. After integrating out the fifth dimension and
diagonalizing mass matrices, we have
\begin{eqnarray}
  {\cal L}_h \;=\; \frac{-1}{v}\sum_n
  \big[(N_d^{\rm t}-\nu_d^{\rm t}V^*)U_N^{\rm t}\big]_{R_n}\!\!
  m_nU_\nu\,\epsilon(\nu_d+VN_d)h^* +\text{h.c.},
\end{eqnarray}
where $[\cdots]_{R_n}$ means the $n$-th mode of the right-handed
component.

\medskip

The heavy neutrino interactions to the SM fields are determined by the
mixing matrix $V$ both in the gauge and Higgs 
vertices. The $3\times\infty$ matrix $V$ is determined by the matrix
forms of neutrino masses in the original 
Lagrangian ${\cal L}+{\cal L}_b$. The matrix elements in $V$ have
the experimental upper bounds from electroweak physics, as will be
seen later. Another important constraint on $V$ comes from the
low-energy neutrino experiments, namely, the seesaw-induced masses
should be of the order of eV scale, which in turn specifies the scale
of heavy neutrino masses $M_N$. This can be seen from the definition
of $V$ by rewriting it with the light and heavy neutrino mass
eigenvalues
\begin{eqnarray}
  V \;=\; (M_\nu^d)^{\frac{1}{2}}P(M_N^d)^{-\frac{1}{2}},
  \label{V}
\end{eqnarray}
where $P$ is an arbitrary $3\times\infty$ matrix 
with $PP^{\rm t}=1$. Therefore one naively expects that, with a fixed
order of $M_\nu^d\sim10^{-1}\,\text{eV}$ and $V\gtrsim10^{-2}$ for the
discovery of experimental signatures of heavy neutrinos, their masses
should be very light and satisfy $M_N^d\lesssim$~keV (this does not
necessarily mean the seesaw operation is not justified as $M_\nu^d$ is
fixed). The previous collider studies on TeV-scale right-handed
neutrinos~\cite{TeVRH} did not impose the seesaw relation~\eqref{V}
and have to rely on some assumptions for suppressing the necessarily
induced masses $M_\nu$. For example, the neutrino mass matrix has
some singular generation structure, otherwise it leads to the
decoupling of seesaw neutrinos from collider physics.

\medskip

A possible scenario for observable heavy neutrinos is to take a
specific value of bulk Majorana masses. Here we assume that bulk Dirac 
masses vanish but it is easy to include them by attaching wavefunction
factors in the following formulas. The equations of motion without
bulk Majorana masses are solved by simple oscillators and the mass
matrices in four-dimensional effective theory are found
\begin{alignat}{2}
  m_n \,\;&=\, \frac{m}{\sqrt{2^{\delta_{n0}}\pi R}}\,,& \qquad\;\;
  M_{R_{mn}} &=\; \delta_{mn}(M_a+M_v), \nonumber \\[.5mm]
  M_{K_{mn}} &=\; \frac{n}{R}\delta_{mn}\,,& \qquad\;\;
  M_{L_{mn}} &=\; \delta_{mn}(M_a-M_v).
\end{alignat}From these,
we find the seesaw-induced mass matrix and the mixing with heavy modes:
\begin{eqnarray}
  M_\nu &=& \frac{1}{2\pi R}\,m^{\rm t}\frac{\pi RX}{\tan(\pi RX)}\,
  \frac{1}{(M_a+M_v)^*}\,m,  \\[1mm]
  \nu &=& U_\nu\nu_d \,+\frac{1}{\sqrt{2\pi R}}\,m^\dagger\bigg[\,
  \,\frac{1}{M_a+M_v}\,\epsilon N_R^{\,0\,*}
  \nonumber \\
  && \qquad\qquad
  +\sum_{n=1}\frac{\sqrt{2}}{X^{2*}\!-(n/R)^2}\,
  \Big[(M_a-M_v)^*\epsilon N_R^{\,n\,*}+
  \frac{n}{R}\,N_L^{\,n}\Big]\,\bigg],
\end{eqnarray}
where $X^2=(M_a+M_v)^*(M_a-M_v)$. The effect of infinitely many
numbers of KK neutrinos appears as the factor $\tan(\pi RX)$. An
interesting case is that (the eigenvalue(s) of) $X$ takes a specific
value $X\simeq\alpha/R$ where $\alpha$ contains half
integers~\cite{DDG}: the seesaw-induced mass matrix $M_\nu$ is
suppressed by the tangent factor (not only by a large Majorana mass
scale), on the other hand, the heavy mode interaction $V$ is
un-suppressed. This fact realizes the situation that right-handed
neutrinos in the seesaw mechanism are observable at sizable rates in
future collider experiments.

\bigskip

\section{Collider Signatures}

\begin{figure}[t]
\begin{center}
\includegraphics[width=5.2cm]{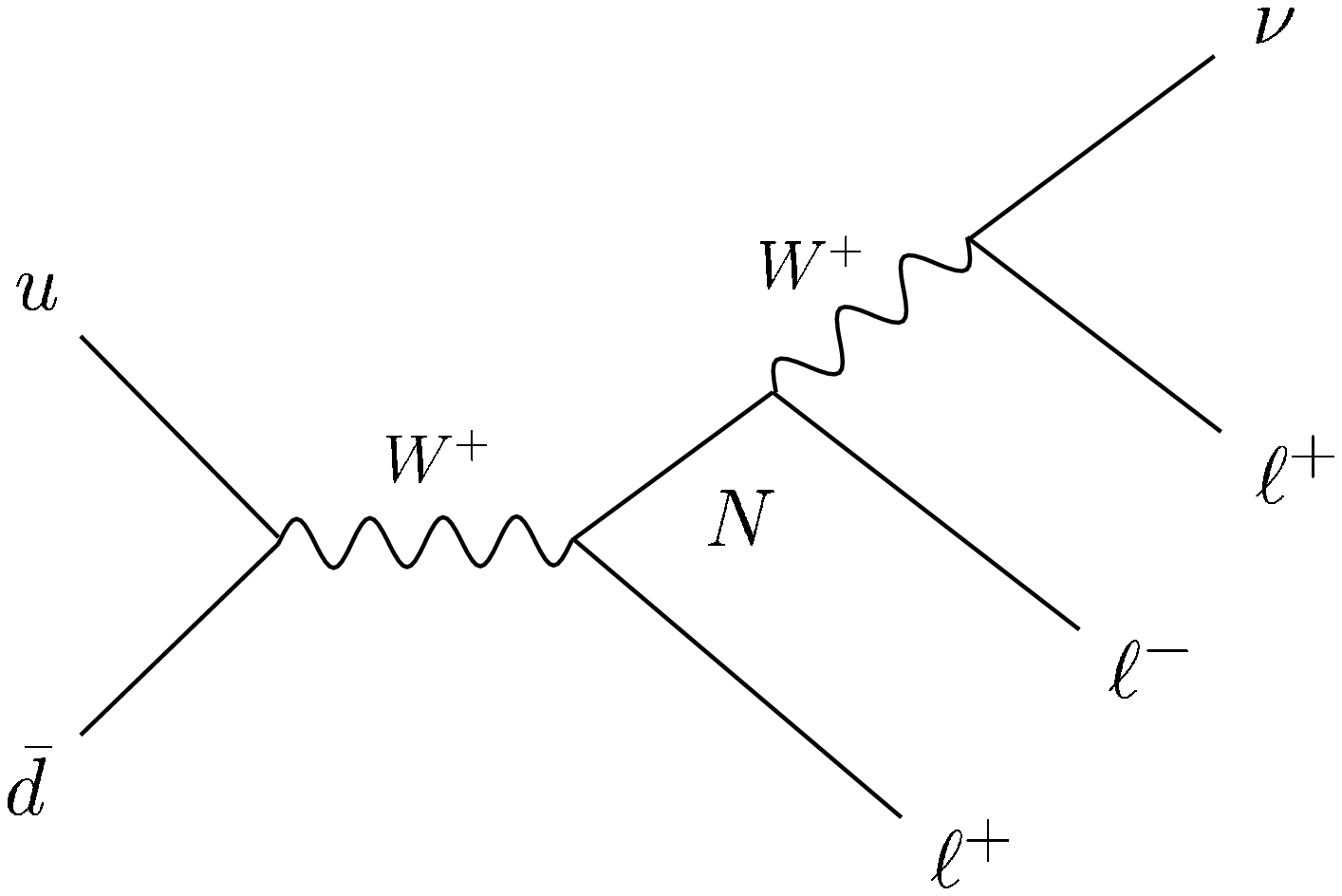}\hspace{15mm}
\includegraphics[width=5.2cm]{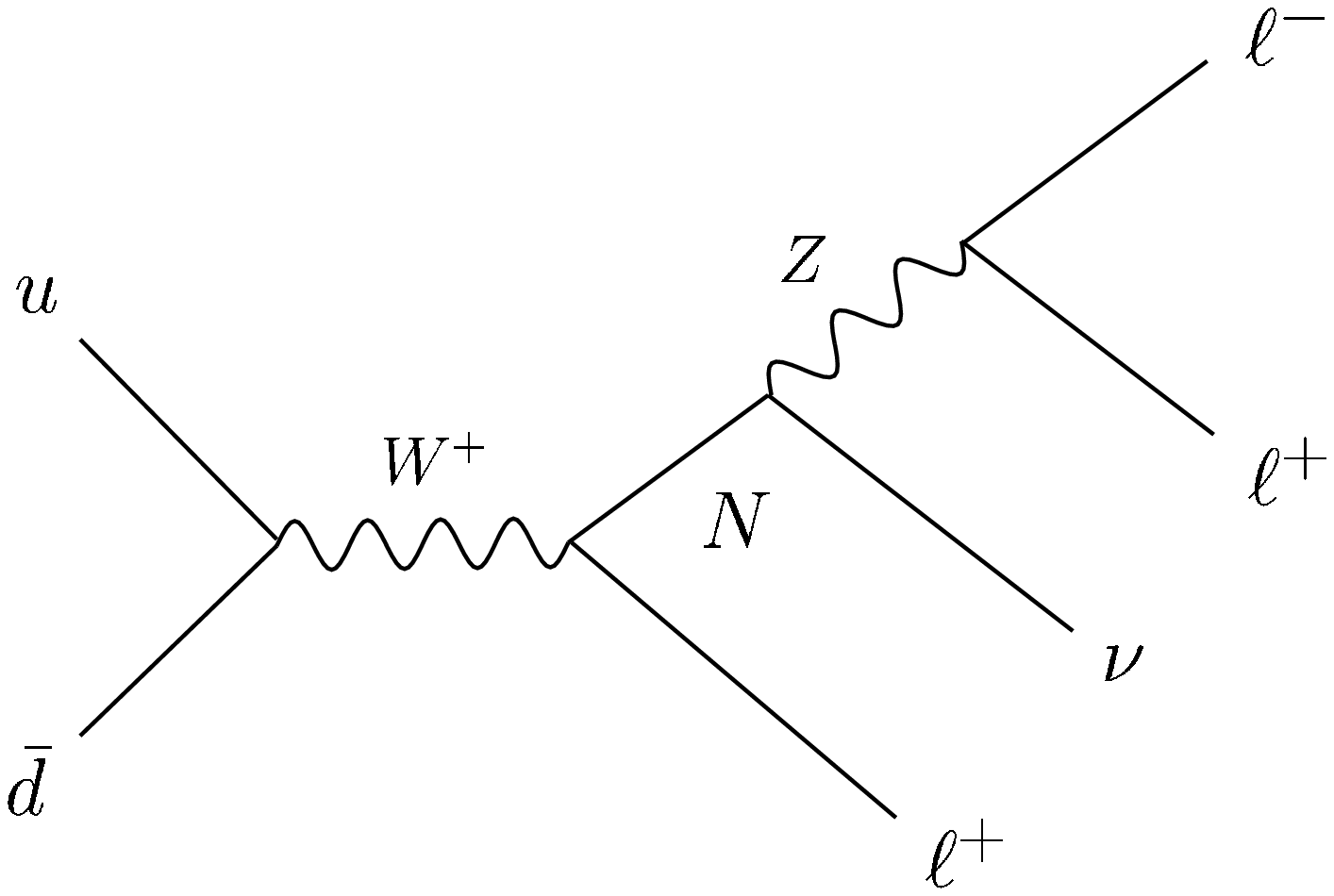}
\caption{Lepton number preserving tri-lepton processes at the LHC.}
\label{fig:3leptons}
\end{center}
\end{figure}

One of the most exciting signals of higher-dimensional theory at
collider experiments is the production of KK excited states. The
signals could be observed at the LHC if new physics, which is
responsible for the generation of neutrino masses, lies around the TeV
scale, and large Yukawa couplings are allowed that lead to a sizable
order of mixing between the left- and right-handed neutrinos. An
immediate question is what processes we should pay attention to find
out the signals. One important possibility is the like-sign di-leptons
signal, $pp\to\ell^+N\to\ell^\pm\ell^\pm W^\mp\to\ell^\pm\ell^\pm jj$, 
because the SM background against the signal is enough
small. Unfortunately, this process violates the lepton number which
should be proportional to tiny Majorana neutrino masses, and is
therefore difficult to be observed at the LHC\@. In this letter we
thus focus on lepton number preserving processes. While there are
various types of such processes related to heavy neutrino productions,
most of these would not be observable due to huge SM backgrounds. As
we will see in the following, an exception suitable for the present
purpose is the tri-lepton signal with large missing transverse 
energy; $pp\to\ell^\pm N\to\ell^\pm\ell^\mp W^\pm\to\ell^\pm\ell^\mp
\ell^\pm\nu(\bar{\nu})$ and $pp\to\ell^\pm N\to
\ell^\pm\nu(\bar{\nu})Z\to\ell^\pm\nu(\bar{\nu})
\ell^\pm\ell^\mp$ (Fig.~\ref{fig:3leptons}). They are possibly
captured at the LHC since only small fractions of SM processes
contribute to the background against the signal.

To investigate the signal quantitatively, we consider the
five-dimensional seesaw theory as a simple example for providing
realistic seesaw neutrino masses and observable collider
signatures. The right-handed Majorana masses 
are $M_a=M$ and $M_v=0$ and diagonalized in the generation space. In
this letter it is assumed that these masses are also generation
independent. As mentioned before, the effective neutrino Majorana
masses become tiny for $M\simeq1/2R$, and thus, the right-handed
neutrino masses can be $M\sim1/R\sim{\cal O}(\text{TeV})$, while
keeping a non-negligible order of Yukawa couplings and sizable
electroweak gauge vertices for the heavy KK neutrinos. We
parametrically introduce a small quantity $\delta$ as
\begin{eqnarray}
  M \;=\; \frac{1-\delta}{2R}.
\end{eqnarray}
Summing up the effects of heavy neutrinos,$\!$\footnote{In theory with
more than one extra dimensions, the sums of infinite KK modes
generally diverse without regularization~\cite{BF}.} we obtain the
seesaw-induced mass $M_\nu=\frac{\delta\pi^2}{8}
\frac{m^{\rm t}m}{M}$. A vanishing value of $\delta$ makes the light
neutrinos exactly massless, where the complete cancellation occurs in
the effects of heavy neutrinos which exhibit the Dirac nature in this
case. The $n$-th excited KK mode spectrum is $M_n=(2n-1)/(2R)$.

The electroweak gauge and Higgs vertices are also evaluated from the
Lagrangian given in the previous section. For example, the neutrino
Yukawa matrix $y$ in the model is expressed as
\begin{eqnarray}
  \frac{y}{\sqrt{2\pi R}} \;=\;
  \frac{2}{\pi v}\frac{1}{\sqrt{\delta R}}\,O^\dagger
  (M_\nu^d)^\frac{1}{2}U^\dagger_{\rm MNS}, 
\end{eqnarray}
where $O$ is the $3\times3$ orthogonal matrix, which generally comes
in reconstructing high-energy quantities from the observable 
ones~\cite{CI}. That corresponds to the matrix $P$ in \eqref{V}. The
model therefore contains the parameters $R$, $\delta$, $M_\nu^d$, 
$U_{\rm MNS}$, and $O$. The neutrino mass differences and the
generation mixing parameters have been measured and we take their
typical experimental values~\cite{analysis}: 
$\Delta m_{21}^2=8\times10^{-5}\>\text{eV}^2$, 
$\,\Delta m_{32}^2=2.5\times 10^{-3}\>\text{eV}^2$, 
$\,\sin\theta_{12}=0.56$, $\,\sin\theta_{23}=0.71$, and 
$|\sin\theta_{13}|\leq0.22$. In this letter we consider the neutrino
mass spectrum with the normal hierarchy. The other cases of the
inverted and degenerate mass patterns can also be analyzed in similar
fashion. The Majorana phases in $U_{\rm MNS}$ have no physical
relevance in the present work and are set to be zero. The remaining
quantities suffer from experimental constraints in low-energy
physics. In particular, the dominant constraint is found to come from
the experimental search for lepton flavor-changing 
processes~\cite{LFV,lowene}. For a real orthogonal matrix $O$, the
limits imposed by lepton flavor conservation are summarized as
\begin{eqnarray}
  \frac{2R}{\delta}\,U_{\rm MNS}\,M_\nu\,U_{\rm MNS}^\dagger
  \;\,\leq\; \left(\begin{array}{ccc}
    10^{-2} & 7\times 10^{-5} & 1.6\times 10^{-2} \\
    7\times 10^{-5} & 10^{-2} & 10^{-2} \\
    1.6\times 10^{-2} & 10^{-2} & 10^{-2}  
  \end{array}\right),
  \label{LFVexp}
\end{eqnarray}
which shows that the most severe limit is given by the 1-2 component,
i.e.\ the $\mu\to e\gamma$ search. We fix $\sin\theta_{13}=0.07$ as a
typical example, and accordingly the Dirac CP phase 
in $U_{\rm MNS}$ is $\phi_D=\pi$ such that the effect of lepton flavor
violation is minimized. It then turns out from \eqref{LFVexp} that all
the constraints are satisfied 
for $\delta/R\geq6.6\,\text{eV}$\@. Finally, the SM Higgs mass is to
be $m_h=120$ GeV in evaluating the decay widths of heavy KK 
neutrinos ($N\to h+\nu$).

\begin{figure}[t]
\begin{center}
\includegraphics[height=7cm]{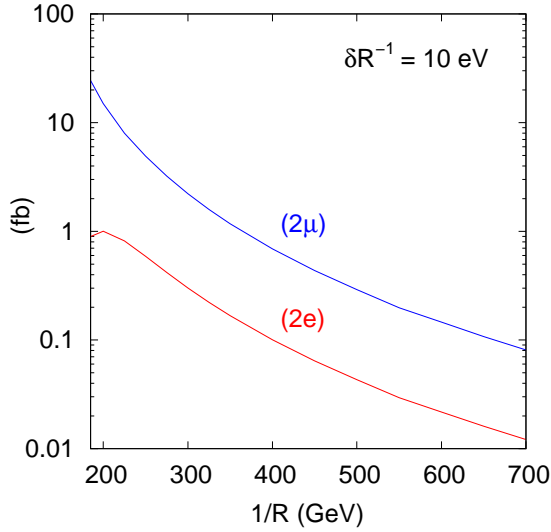}
\caption{Total cross sections of tri-lepton signals as the functions
of the compactification scale $R$ with a fixed value $\delta/R=10$ eV.}
\label{fig:Xsection}
\end{center}
\end{figure}

\medskip

Now we are at the stage of investigating the tri-lepton signal of 
heavy neutrino productions at the LHC\@. Since the tau lepton is
hardly detected compared to the others, we consider the signal event
including only electrons and muons. There are four kinds of tri-lepton
signals: $eee$, $ee\mu$, $e\mu\mu$, and $\mu\mu\mu$. In this work, we
use two combined signals which are composed 
of $eee+ee\mu$ (the $2e$ signal) and $e\mu\mu+
\mu\mu\mu$ (the $2\mu$ signal). Figure~\ref{fig:Xsection} shows the
total cross sections for these signals from the 1st KK mode
productions at the LHC\@. They are described as the functions of the
compactification scale $R$ with $\delta/R$ being 10 eV\@. It is found
from the figure that the cross section for the $2\mu$ signal is about
one order of magnitude larger than the $2e$ signal.\footnote{For the
inverted hierarchy spectrum of light neutrinos, the $2e$ signal cross
section becomes larger than the $2\mu$ one.} We have also evaluated
the cross sections of tri-lepton signals from heavier KK neutrinos and
found that they are more than one order of magnitude smaller than the
above and are out of reach of the LHC machine. A high luminosity
collider with clean environment such as the International Linear
Collider (ILC) would distinctly discover the signatures of KK mode
resonances.

\begin{figure}[t]
\begin{center}
\includegraphics[height=7cm]{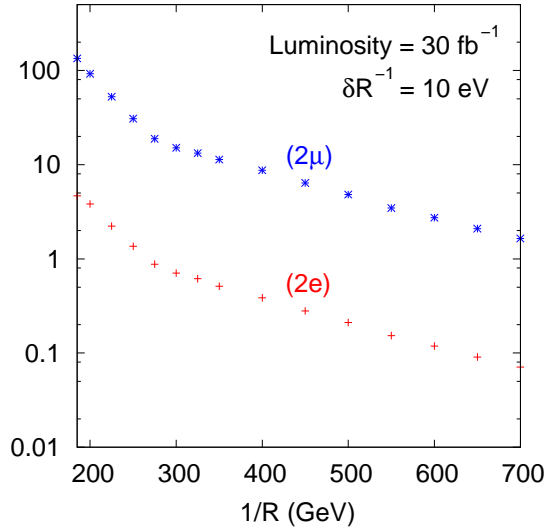}
\caption{Expected event numbers of the $2e$ and $2\mu$ signals after
implementing the kinematical cut. The event numbers are depicted as
the functions of the compactification scale $R$ with a fixed 
value $\delta/R=10$ eV\@. The integrated luminosity is taken to 
be 30 fb$^{-1}$.}
\label{fig:signal}
\end{center}
\end{figure}

To clarify whether the tri-lepton signal is captured at the LHC, it is
important to estimate SM backgrounds against the signal. The SM
backgrounds which produce or mimic the tri-leptons final state have
been studied~\cite{cut,cut2}, and for the present purpose a useful
kinematical cut is discussed to reduce these SM
processes~\cite{cut2}. According to that work, we adopt the following
kinematical cuts;
\begin{itemize}
\setlength{\parskip}{0pt}
\item The existence of two like-sign charged 
leptons $\ell_1^\pm$, $\ell_2^\pm$, and an additional one with the
opposite charge $\ell_3^\mp$.
\item Both energies of the like-sign leptons are larger than 30 GeV.
\item Both invariant masses 
from $\ell_1$ and $\ell_3$ and from $\ell_2$ and $\ell_3$ are 
larger than $m_Z+10$ GeV or smaller than $m_Z-10$ GeV.
\end{itemize}
The last condition is imposed to reduce large backgrounds from the
leptonic decays of $Z$ bosons in the SM
processes. Figure~\ref{fig:signal} shows the expected numbers of
signal events after imposing the cuts stated above. The results are
depicted by assuming the integrated luminosity 30 fb$^{-1}$. In order
to estimate the efficiency for signal events due to the cuts, we have
used the Monte Carlo simulation using the CalcHep
code~\cite{CalcHep}. Since the event numbers of SM backgrounds after
the cut are about 260 for the $2e$ signal and 110 for 
the $2\mu$ signal~\cite{cut2}, the $2\mu$ events are expected to be 
observed if $1/R$ is less than a few hundred GeV.

\begin{figure}[t]
\begin{center}
\includegraphics[height=7cm]{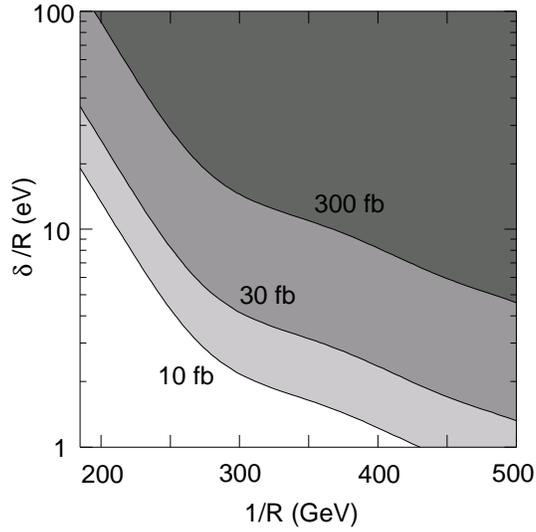}
\caption{Luminosity for the 3$\sigma$ reach on 
the $(1/R,\delta/R)$ plane (10, 30, and 300 fb$^{-1}$ contours).} 
\label{fig:Reach2mN}
\end{center}
\end{figure}

The luminosity which is required to find the 2$\mu$ signal at the LHC
is shown in Fig.~\ref{fig:Reach2mN} as a contour plot on 
the $(1/R,\delta/R)$ plane. The contour is obtained by computing the
significance for the signal discovery,
\begin{eqnarray}
  S_{ig} \;\equiv\; \frac{S}{\sqrt{S+B}},
\end{eqnarray}
where $S$ and $B$ are the numbers of the 2$\mu$ events and the
corresponding SM backgrounds after the kinematical cut. Since
both $S$ and $B$ are proportional to the luminosity, it is possible to
estimate the luminosity, e.g.\ giving $S_{ig}=3$ which is plotted in
Fig.~\ref{fig:Reach2mN}. The luminosity for signal 
confirmation (for $S_{ig}=5$) are also obtained by scaling the above
result. The luminosity of 10, 30, and 300 fb$^{-1}$ are depicted in
the figure. It is found that if $1/R$ is less than about 250 GeV, the
signals will be observed at the early run of the LHC, while a larger
luminosity is needed for a smaller size of extra dimension to find its
signals.

\bigskip

\section{Summary and Discussion}

We have discussed a seesaw scenario where right-handed neutrinos are
around TeV scale, accessible in near future particle experiments. The
seesaw-induced mass scale is of the order of eV, while the
right-handed neutrinos have sizable gauge and Yukawa couplings to the
SM sector. The scenario is a five-dimensional extension of the SM
with right-handed neutrinos, where the ordinary SM particles locally
live in four dimensions and the right-handed neutrinos exist in the
bulk. The light neutrinos obtain tiny Majorana masses due to the small 
lepton number violation, and therefore the same-sign di-lepton
processes cannot be observed. We have analyzed, as the most effective
LHC signal, the lepton number preserving processes with tri-lepton
final states, $pp\to\ell^\pm\ell^\pm\ell^\mp\nu(\bar\nu)$. It is found 
that the scenario gives enough excessive tri-lepton events beyond the
SM backgrounds in wide regions of parameter space, and the LHC would
discover the signs of tiny neutrino mass generation and extra
dimensions.

The possible experimental detections of neutrino mass generations have
been discussed in other seesaw
scenarios~\cite{dileptons,TeVRH,others}. In the present analysis, only
the 1st excited mode contributes to the signals. The observation of
higher KK modes is expected to be within the reach of future collider
experiments such as the ILC, which result makes the scenario
substantially confirmed. Further analysis of such collider signatures,
together with including bulk Dirac masses and curved gravitational
backgrounds, are left for important future study.

\bigskip
\subsection*{Acknowledgments}
\noindent
The authors thank Takahiro Nishinaka for collaboration during the
early stage of this work. This work is supported in part by the
scientific grant from the ministry of education, science, sports, and
culture of Japan (Nos. 20540272, 20039006, 20244028, 18204024,
20025004, and 20740135), and also by the grant-in-aid for the global
COE program "The next generation of physics, spun from universality
and emergence" and the grant-in-aid for the scientific research on
priority area (\#441) "Progress in elementary particle physics of the
21st century through discoveries of Higgs boson and supersymmetry"
(No.~16081209).

\newpage

\end{document}